\documentclass[preprint,12pt]{elsarticle}

\usepackage{enumitem}
\usepackage{color}
\usepackage{amssymb}
\usepackage{graphicx}
\usepackage{booktabs}
\usepackage{hyperref} 
\usepackage{multirow}

\journal{Artificial Intelligence in Medicine}

\begin{document}

\begin{frontmatter}



\title{Uncovering the Genetic Basis of Glioblastoma Heterogeneity through Multimodal Analysis of Whole Slide Images and RNA Sequencing Data}


\author[inst1,inst3]{Ahmad Berjaoui\fnref{note1}\corref{note2}}
\author[inst1]{Eduardo Hugo Sanchez\fnref{note1}}
\author[inst1]{Louis Roussel}
\author[inst2,inst3]{Elizabeth Cohen-Jonathan Moyal}

\affiliation[inst1]{organization={IRT Saint Exupéry},
            city={Toulouse},
            country={France}}

\affiliation[inst2]{organization={Oncopole Claudius Regaud},
            city={Toulouse},
            country={France}}

\affiliation[inst3]{
organization={INSERM UMR 1037 Cancer Research Center of Toulouse (CRCT), University Paul Sabatier Toulouse III},
            city={Toulouse},
            country={France}
}

\fntext[note1]{Equal contribution}

\cortext[note2]{Corresponding author}

\begin{abstract}
Glioblastoma is a highly aggressive form of brain cancer characterized by rapid progression and poor prognosis. Despite advances in treatment, the underlying genetic mechanisms driving this aggressiveness remain poorly understood. In this study, we employed multimodal deep learning approaches to investigate glioblastoma heterogeneity using joint image/RNA-seq analysis. Our results reveal novel genes associated with glioblastoma. By leveraging a combination of whole-slide images and RNA-seq, as well as introducing novel methods to encode RNA-seq data, we identified specific genetic profiles that may explain different patterns of glioblastoma progression. These findings provide new insights into the genetic mechanisms underlying glioblastoma heterogeneity and highlight potential targets for therapeutic intervention. Code and data downloading instructions are available at: \url{https://github.com/ma3oun/gbheterogeneity}.
\end{abstract}



\end{frontmatter}


\section{Introduction}
\label{sec:intro}
Glioblastoma (GB) is the most aggressive primary brain tumor and is not curable \cite{taylor_glioblastoma_2019} despite of treatment associating surgery when possible followed by radio-chemotherapy \cite{stupp_radiotherapy_2005} and more recently Tumor treating Fields \cite{rominiyi2021tumour} leading to a median overall survival (OS) of 20.9 months and a progression free survival (PFS) around 7 months. Despite being a minor population of cancer cells, the cancer stem cells that are identified in glioblastoma (GSCs) are thought to be the major driving force behind glioblastoma biological heterogeneity and are likely to explain the high rates of glioblastoma recurrence. In the STEMRI clinical trial aiming to study GB heterogeneity and the enrichment of GSC in certain areas defined by multimodal Magnetic Resonance Imaging (MRI) (NCT01872221) \cite{lemarie2023stemri} different GSC sub-populations extracted from tumor samples obtained by multimodal MRI guided surgery were xenografted into mice brain to study their invasion patterns as well as their aggressiveness. RNA-sequencing (RNA-seq) on each tumor bulk samples was also performed. The observed differences in mice survival according to the GSC implanted confirm the heterogeneous nature of tumor cells lineage.\\
In this study, we set out to determine potential genetic markers associated with glioblastoma aggressiveness using multimodal deep learning. Our results reveal genetic targets already identified in medical literature but also highlight new potential targets. We leverage the extensive recent developments in multimodal data analysis, namely image and text, and adapt these techniques to whole slide images (WSI) and RNA-seq data. Training uses public data from The Cancer Genome Atlas (TCGA) \cite{collins2007mapping} but also WSI and RNA-seq data from the STEMRI trial.
Furthermore, we introduce a new RNA-seq encoding technique where genes are grouped based on biological pathways prior to encoding and show better performance in comparison to mere position based grouping.\\
Overall, our results can be used to test new GB treatment strategies. Our main contributions can be summarized as follows:
\begin{enumerate}[label=\roman*]
\item Identify genetic profiles leading to unique GB patterns.
\item Novel method to encode RNA-seq data for use in deep learning models.
\item Novel method to combine WSI and RNA-seq data for use in deep learning models.
\end{enumerate}

\paragraph{Related work} This study pioneers the application of artificial intelligence (AI) algorithms to investigate the heterogeneous nature of glioblastoma, diverging from the common focus on specific biomarkers. Extensive research has been conducted utilizing AI algorithms on MRI data for precise tumor grading, OS prediction, and biomarker identification, as evidenced by studies such as \cite{henssen2023challenges, zlochower2020deep, pouessel2023hypofractionated, robinet2023mri}. Furthermore, recent investigations have explored the potential of integrating multimodal data sources, including MRI, WSI, and RNA-seq, to predict OS or enhance tumor image segmentation \cite{alleman2023multimodal,fathi2022clinical,robinet2024drim}. However, these studies adopt a global approach to the disease, overlooking its heterogeneous characteristics.\\

This paper is structured as follows. Section \ref{sec:mat} details the developed algorithms, the datasets and training strategies. Section \ref{sec:exp} showcases our experimental results followed by a discussion of our main findings in section \ref{sec:disc}.

\section{Material and methods}
\label{sec:mat}

\subsection{Datasets}
\label{subsec:dataset}
\begin{figure}[htbp]
    \centering
    \includegraphics[width=0.49\linewidth]{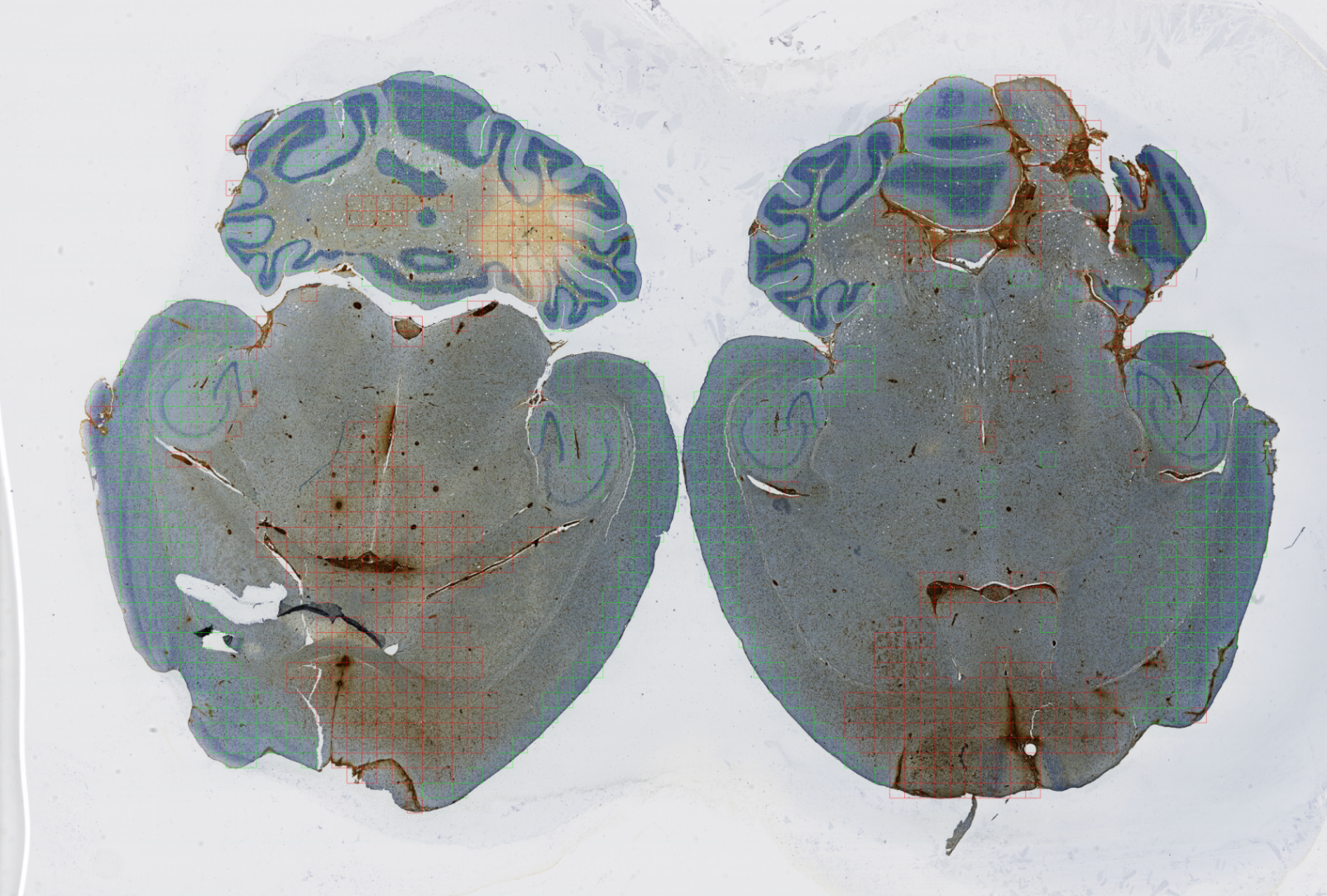}
    \includegraphics[width=0.49\linewidth]{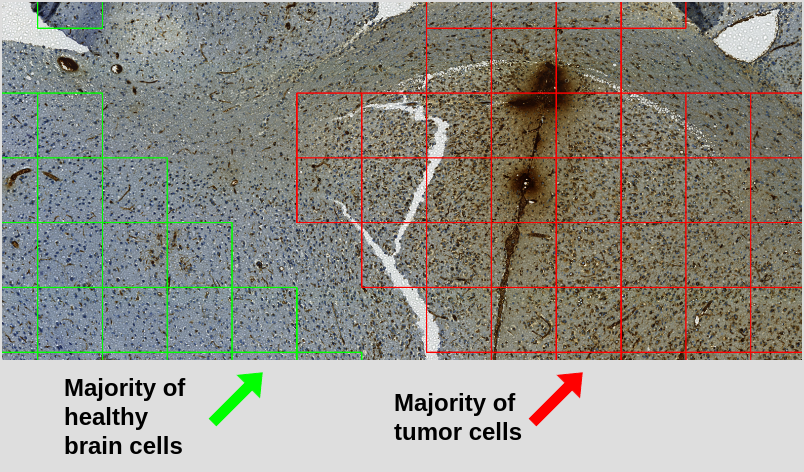}
    \caption{(Left) Example of a WSI of a mouse brain slice. (Right) Zoom on the upper-right part of the left brain slice. Red square patches have a majority of tumor cells whereas green square patches have a majority of non-tumor cells.}
    \label{fig:wsi}
\end{figure}

Two datasets are used to train our models. The TCGA public dataset is used to pre-train RNA-seq encoders. This dataset contains around 10k RNA-seq gene expression samples from multiple sites (breast, brain, prostate, bladder, etc.).\\
51 RNA-seq samples from the STEMRI trial complete this data. These correspond to 16 patients and bulk tumor RNA from metabolically heterogeneous regions identified by spectral MRI, as explained in \cite{lemarie2023stemri}. This corresponds to 51 different tumor cell lineages. To insure data homogeneity, only genes common to both STEMRI and TCGA were kept. Genes expressions with relatively low variance were also removed and all data was then normalized.\\
Tumor cells from these 51 different lineages were xenografted onto mice brains. Cells from the same lineage were used onto more than one mouse whenever possible, as culture was not always successful. This led to a total of 116 mice. Whole slide images of mice brain slices were used at x10 magnification factor (fig.\ref{fig:wsi}) after each mouse's death. These images are in average composed of 16k x 21k RGB pixels and human tumor cells are highlighted using specific coloration.\\
Observation data contains patient OS and PFS and mouse survival time in days.

\subsection{RNA-seq encoding}
\label{subsec:rna}
\begin{figure}[htbp]
    \centering
    \includegraphics[width=0.95\linewidth]{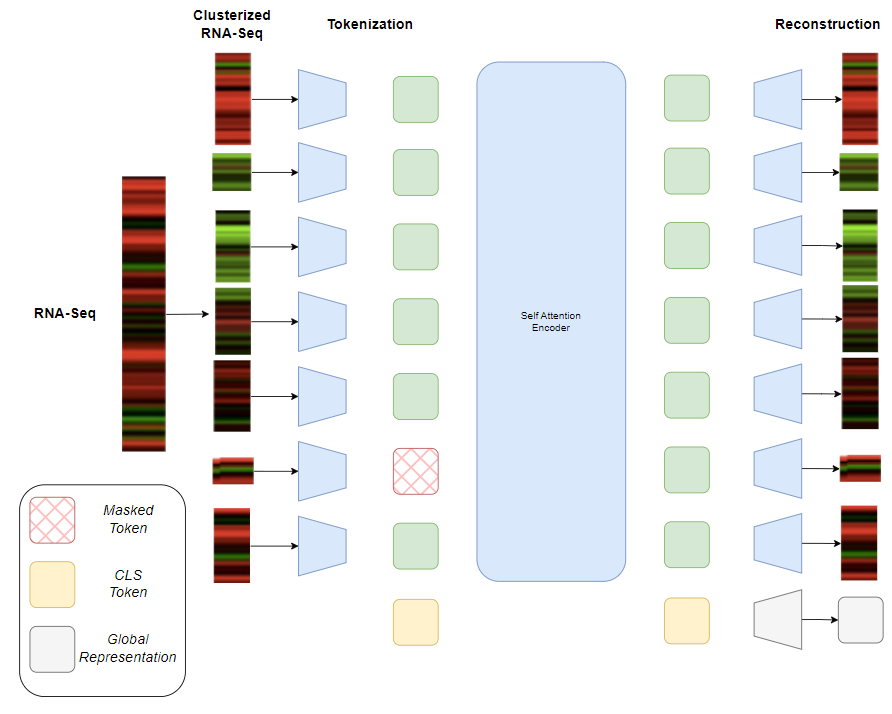}
    \caption{RNA-seq encoding. The original vector is reorganized according to directed graph clustering from the PPI knowledge graph. Sub-vectors are projected to a common embedding size. A token is randomly masked. The decoder reconstructs the RNA-seq vector using the remaining tokens and the \textit{cls} token.}
    \label{fig:rna}
\end{figure}

RNA-seq data is a vector representing the expressions for roughly 19K protein coding genes.  Moreover, attention-based encoders are currently state-of-the-art for vision \cite{dosovitskiy2020image} and language \cite{radford2019language}. Combining these two observations, we leveraged the Protein-to-Protein interaction  (PPI) graph \cite{chandak2023building} to regroup protein scores before computing attention scores. The PPI graph is a directed graph and \cite{dugue2022direction} propose a clustering algorithm that considers the directed nature of PPI.\\
The reader can refer to \cite{dugue2022direction} for more detail  about the directed Louvain algorithm but in brief, the aim is to maximize \textit{modularity} \cite{newman2003structure} $Q_d$ which in the case of a directed graph, can be defined by:
\begin{equation}
    Q_{d} = \frac{1}{m}\sum_{i,j}[A_{ij}-\frac{d^{in}_i d^{out}_j}{m}]\delta(c_i,c_j)
\end{equation}
$m$ is the total number of edges. $A_{ij}$ indicates the presence of an edge between nodes $i$ and $j$. $d^{in}_i$ (\textit{resp.} $d^{out}_i$) is the in-degree (\textit{resp.} out-degree) of node $i$. The degree is the number of incoming (\textit{resp.} outgoing) edges. $\delta(c_i,c_j)$ indicates whether nodes $i$ and $j$ are in the same cluster. Connected edges with low degrees contribute significantly to modularity when they are in the same cluster as this is much less likely than a connection between edges of high degrees, regardless of cluster assignment.\\
Applying this algorithm to PPI leads to 11 clusters. Clusters with significantly less nodes were then grouped together, leading to 7 final clusters with an average of 2905 genes per cluster (ranging between 1514 and 3182).\\
Genes in each cluster constitute a sub-vector that is projected to a common size. The result is 7 sub-vectors of the same size, each as a separate token as input to a masked-autoencoder \cite{he2022masked}. During training, one token is randomly masked and the decoder uses the \textit{cls} token and the remaining tokens to reconstruct it, as illustrated in fig.\ref{fig:rna}. This results in a first loss term:
\begin{equation}
    \mathcal{L}^r_{RNA}=\frac{1}{N}\sum_{i=1}^N ||\mathbf{x}_i -\mathbf{\hat{x_i}}||^2    
    \label{eq:rna_r}
\end{equation}
where $N$ is the batch size, $\mathbf{x}_i$ represents RNA-seq vector $i$ and $\mathbf{\hat{x_i}}$ its reconstruction.
In our experiments, we compared this encoding strategy to a similar strategy but where genes where clustered according to their chromosome. The second loss term $\mathcal{L}^c_{RNA}$ is obtained using pairwise contrastive learning \cite{chen2020simple}: a positive pair is obtained from two RNA sub-vectors corresponding to the same tumor cell lineage, and a negative pair is obtained using two sub-vectors corresponding to two different tumor cell lineages. This forces the encoder to learn lineage specific features. Given a pair of representations $\mathbf{z}_i$ and $\mathbf{z}'_i$ with batch size $N$, temperature $\tau$, the loss term $l_i$ is therefore: 
\begin{equation}
    l_i = -y(i,i')\log\frac{\exp(\mathbf{z}_i^{T}\mathbf{z}'_i/\tau)}{\sum_{j=1}^{N}\exp(\mathbf{z}_i^T\mathbf{z}_j/\tau)}
\end{equation}
$y(i,i')=1$ for a matching pair and $0$ for negative pairs (i.e. different cell lineages).
Hence,
\begin{equation}
    \mathcal{L}^c_{RNA} = \sum_{i=1}^{N}l_i
    \label{eq:rna_c}
\end{equation}
The final loss for RNA-seq encoding is:
\begin{equation}
    \mathcal{L}_{RNA}=\mathcal{L}^r_{RNA} + \mathcal{L}^c_{RNA}
    \label{eq:rna_full}
\end{equation}
We set $\tau=0.005$, a relatively low value, in order to emphasize the difference between dissimilar samples.
\subsection{WSI encoding}
\label{subsec:wsi}
All images are split into sequential, non-overlapping patches of 256x256 RGB pixels, which results in roughly 5K patches per image and a total of 100K patches for the whole dataset. All patches are converted to hue, saturation, luminance, space (HSL). HSL conversion makes it easier to distinguish non-tumor and tumor cells (colored in brown). Patches with a majority of tumor cells can hence be easily extracted by computing an overall pixel score, by counting pixels that respect a given hue interval, regardless of lighting conditions. Patches can then be sorted based on that score, which correlates with their tumor cells content. Only those with a brown colored pixel ratio exceeding 20\% are kept for training, test and validation. The dataset is then split into 90\% for training and 10\% for validation and test.\\

Training a masked autoencoder for tumor patches would require a very high computational cost due to the size of the WSI dataset. We therefore rely solely on contrastive learning as our experiments have proved that it was enough to achieve very good performance. A $256\times256$ patch is split into smaller $16\times16$ sub-patches and we use a ViT \cite{dosovitskiy2020image} transformer encoder to obtain a global representation of the patch, using the \textit{cls} token. This representation serves then as an anchor in a triplet loss\cite{Dong_2018_ECCV}: the anchor ($\mathbf{z}_i$) is matched with the representation of another patch of the same tumor cell lineage ($\mathbf{z}_+$) and it is contrasted with the representation of a patch from another tumor cell lineage ($\mathbf{z}_-$).
\begin{equation}
    \mathcal{L}_{WSI}^{c} = \frac{1}{N}\sum_{i=1}^N\max(||\mathbf{z}_i - \mathbf{z}_+||^2 - ||\mathbf{z}_i - \mathbf{z}_-||^2 + d,0)
    \label{eq:wsi}
\end{equation}
where $d$ is the margin and $N$ is the batch size. We set $d=1\times10^{-4}$ after a coarse grid-search. For each positive pair, we select the negative sample that leads to the greatest triplet loss value.

\subsection{Multimodal training}
\label{subsec:mulimodal}
Both the RNA and the WSI encoders are pre-trained using eq.\ref{eq:rna_full} and eq.\ref{eq:wsi}. In order to consider both modalities, we draw inspiration from ALBEF\cite{li2021align}. A multimodal contrastive loss $\mathcal{L}_{MM}^c$ is used to align the modalities' representations before a cross-attention encoder. For a given RNA vector and a 256x256 WSI patch, let $\mathbf{u}_k$ and $\mathbf{u}'_k$ be the \textit{cls} representations at the output of their respective encoders, after linear projection to a common embedding size. Order is irrelevant as the following equations treat $\mathbf{u}_k$ and $\mathbf{u}'_k$ symmetrically. We use a similar pairwise loss as in eq.\ref{eq:rna_c}, considering a pair of RNA/WSI representations $\mathbf{u}_k$ and $\mathbf{u}'_k$, a temperature $\tau_c$ and batch size $N$:
\begin{equation}
    l_k = -y(k,k')\log\frac{\exp(\mathbf{u}_k^T\mathbf{u}'_k/\tau_c)}{\sum_{j=1}^{N}\exp(\mathbf{u}_k^T\mathbf{u}_j/\tau_c)}
\end{equation}
$y(k,k')=1$ for a matching pair and $0$ for negative pairs.
\begin{equation}
    \mathcal{L}_{MM}^c = \sum_{k=1}^{N}l_{k}
\end{equation}
\begin{figure}[htbp]
    \centering
    \includegraphics[width=0.95\linewidth]{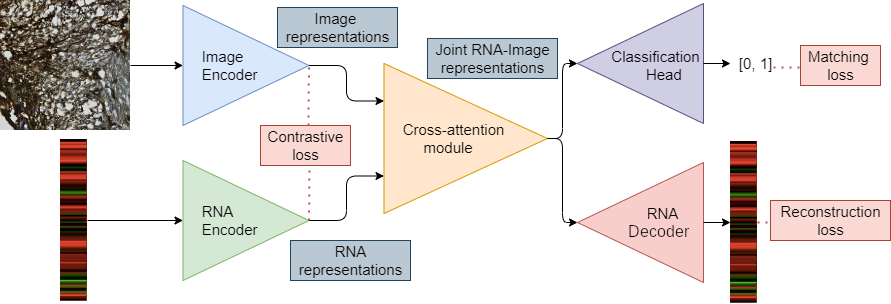}
    \caption{Multimodal model. A pairwise contrastive loss aligns RNA and WSI representations. A cross-attention module is used to obtain a joint representation. The latter is used for a classification head that matches data from the same tumor cells lineage and a RNA decoder reconstructs the original RNA vector.}
    \label{fig:multimodal}
\end{figure}
As is done in \cite{li2021align}, a cross-attention model is used to obtain a joint RNA/WSI representation from the unimodal representations. The cross-attention model is trained along with a classification head and a RNA decoder head. The classification head is trained using a weakly supervised loss $\mathcal{L}_{MM}^{m}$ where a label of 1 indicates samples from the same tumor cells lineage and 0 for mismatching samples. The RNA decoder is trained using a reconstruction loss $\mathcal{L}_{MM}^{r}$ similar to Eq.\ref{eq:rna_r}.

\subsection{Evaluation}
\label{subsec:eval}
\begin{figure}[htbp]
    \centering
    \includegraphics[width=0.95\linewidth]{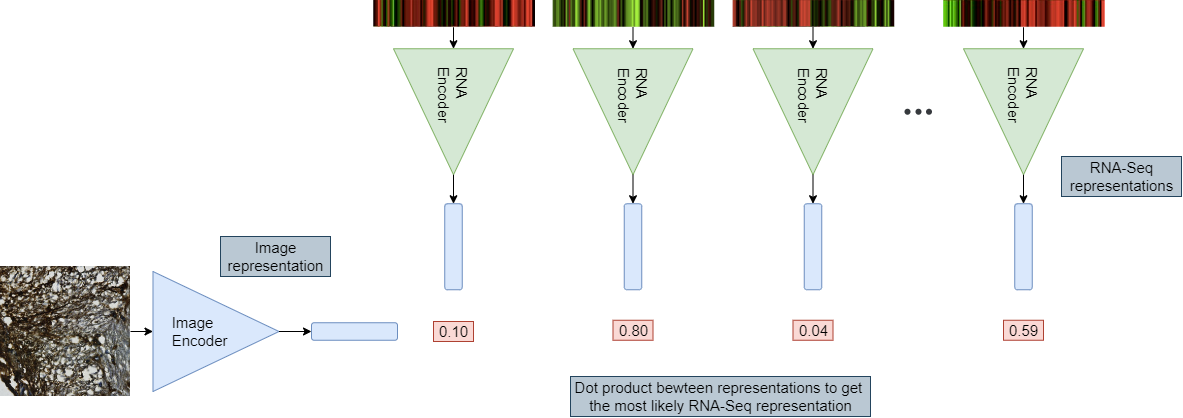}
    \caption{RNA-seq retrieval using a WSI patch.}
    \label{fig:retrieval}
\end{figure}

Training the multimodal model leads to aligned WSI and RNA-seq representations. Given a WSI patch not used during training, it becomes possible to retrieve the closest RNA-seq vector amongst all the 51 RNA-seq vectors by measuring its cosine similarity, as shown in fig.\ref{fig:retrieval}

The main evaluation criteria is therefore matching accuracy, computed over a subset of WSI patches (from a total of 100K patches) not used during training but that are drawn evenly amongst corresponding cell lineages and patients. Let $\mathcal{I}=\{1..51\}$ be the set of all the 51 RNA-seq samples, $\mathbf{z}_{WSI}$ be the output of the image encoder for a given WSI patch, and $\mathbf{u}_i$ be the output of the RNA encoder for sample $i$ ($i\in\mathcal{I}$). The predicted sample $\hat{p}$ is given by eq.\ref{eq:retrieval}
\begin{equation}
    \hat{p} = \arg\max_{i\in\mathcal{I}}\frac{\mathbf{z}^T_{WSI}\mathbf{u}_i}{||\mathbf{z}_{WSI}||\cdot||\mathbf{u}_i||}
    \label{eq:retrieval}
\end{equation}

\section{Experiments}
\label{sec:exp}
\begin{figure}[htbp]
    \centering
    \includegraphics[width=0.8\linewidth]{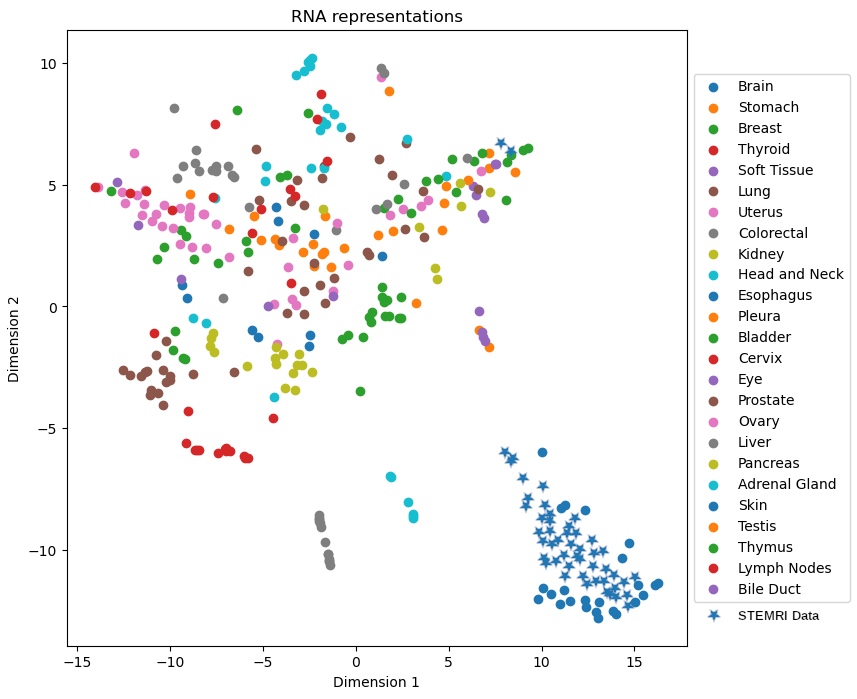}
    \caption{2D t-SNE projections of RNA-seq representations of primary tumor locations, combining public TCGA data (circles) and STEMRI data (stars). The figure clearly showcases a strong similarity between public brain RNA-seq test data and STEMRI data.}
    \label{fig:rna-tsne}
\end{figure}

\paragraph{RNA-encoding}
After pre-training the RNA-seq encoder depicted in §\ref{subsec:rna} using TCGA training data, we plotted the t-SNE two dimensional projections for both TCGA test data and STEMRI data (fig.\ref{fig:rna-tsne}). Nearly all STEMRI data points lie very close to TCGA data points corresponding to brain tumors.
\label{subsec:exp-rna}
\paragraph{RNA retrieval} Table.\ref{tab:results} lists RNA-seq retrieval accuracy considering tumor cell lineage and patient of origin. Accuracy is computed using WSI patches from the test set.  The best results are obtained using pathway based clustering as explained in §\ref{subsec:rna}, and using only a matching loss (without a RNA decoder and a reconstruction loss). Results show a near perfect accuracy for both patient and cell lineage matching tasks. This indicates that our model has learned unique genetic features that can be matched against unique cellular microscopic patterns. The table also shows good matching accuracy in the configuration where there is no matching loss (86.1\% for lineage matching), serving as further proof of the high quality of RNA-seq and WSI representations, regardless of the limited number of patients and cell lineages. This means that the trained WSI and RNA-seq encoders can generalize well to unseen independent data.
\begin{table}[htbp]
    \centering
    \resizebox{0.99\textwidth}{!}{
\begin{tabular}{lcc}
\hline
Model & Accuracy (patient) &  Accuracy (lineage)  \\
\hline
Random initialization                                        & 0.050 & 0.006  \\
$\mathcal{L}_{MM}^{r} + \mathcal{L}_{MM}^{c} + \mathcal{L}_{MM}^{m}$ (frozen RNA-Image encoders)          & 0.29 & 0.199 \\
$\mathcal{L}_{MM}^{r} + \mathcal{L}_{MM}^{c}$ (all modules are trainable)          & 0.874 & 0.861 \\
$\mathcal{L}_{MM}^{r} + \mathcal{L}_{MM}^{c} + \mathcal{L}_{MM}^{m}$ (all modules are trainable)          & 0.946 & 0.941 \\
$\mathcal{L}_{MM}^{c} + \mathcal{L}_{MM}^{m}$ (all modules are trainable/chromosome gene groups) & 0.890 & 0.913 \\ 
$\mathcal{L}_{MM}^{c} + \mathcal{L}_{MM}^{m}$ (all modules are trainable/pathways gene groups) & \textbf{0.970} & \textbf{0.968} \\ 
\hline

\end{tabular}
}
\caption{Lineage and patient retrieval results. The table lists WSI patch/RNA-seq matching accuracy by considering cell lineage or patient of origin, in several configurations.}
\label{tab:results}
\end{table}

\begin{figure}[htbp]
    \centering
    \includegraphics[width=0.99\linewidth]{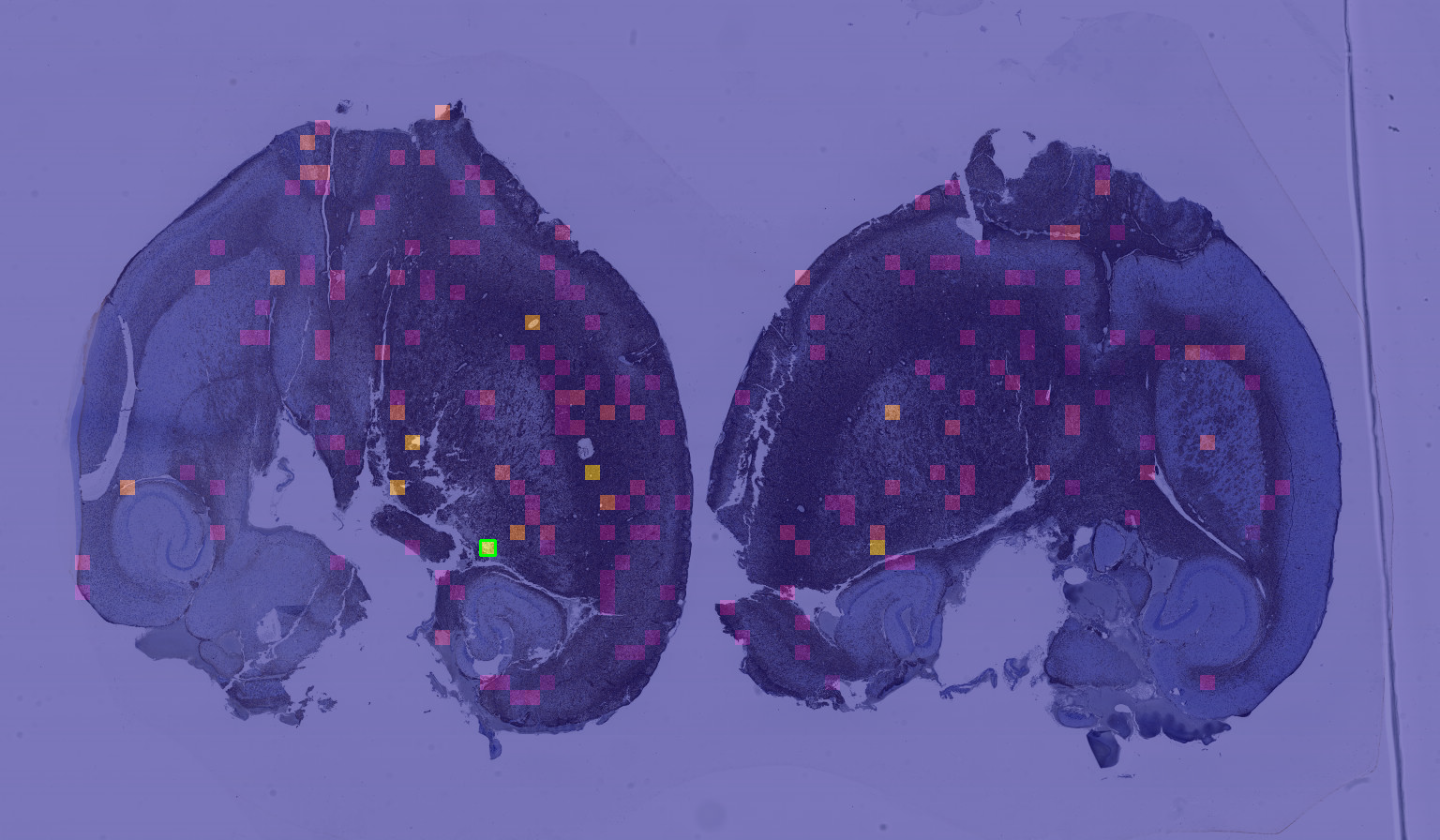}
    \includegraphics[width=0.49\linewidth]{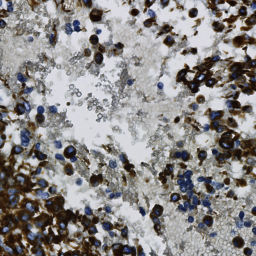}
    \includegraphics[width=0.49\linewidth]{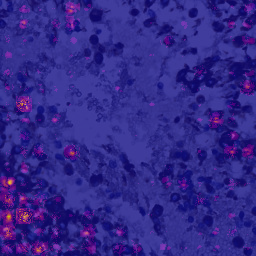}
    \caption{(Upper center) Example of a WSI of a mouse brain slice where patches (only from the test set) have been colored according to how good they match the corresponding RNA-seq. The patch with the highest matching score (green box) is shown in the lower left image. Grad-CAM is applied to that patch in the lower right image.}
    \label{fig:gradcam}
\end{figure}

\paragraph{Genetic analysis} 
Grad-CAM \cite{selvaraju2020grad} is a common technique to analyze deep neural networks' response. We used grad-CAM to determine genetic positions that have the greatest contribution in the WSI/RNA-seq matching process. For each WSI in the test set, we sort gene expressions according to their importance as computed using grad-CAM. Keeping the 15 most occurring gene expressions leads to the results in table.\ref{tab:genes}. \\
Grad-CAM has also been applied to WSI patches, as illustrated in fig.\ref{fig:gradcam}, where it highlights tumor cells invading healthy tissue. This suggests that the model focuses on invasion patterns to determine the best RNA-seq match, offering a valuable tool for analyzing tumor behavior and identifying potential biomarkers.

\begin{table}[htbp]
    \centering
    \resizebox{0.70\textwidth}{!}{
\begin{tabular}{llrl}
\toprule
{} &                \textbf{Gene} &  \textbf{Occurrence} &   \textbf{Symbol} \\
\hline\hline
0  &  ENSG00000198964.14 &       2344 &    \colorbox{cyan}{SGMS1} \\
1  &  ENSG00000120008.16 &       2151 &    \colorbox{cyan}{WDR11} \\
2  &  ENSG00000111670.16 &       1759 &   GNPTAB \\
3   &  ENSG00000078098.14 &       1669 &      \colorbox{cyan}{FAP} \\
4  &  ENSG00000143147.14 &       1639 &   \colorbox{cyan}{GPR161} \\
5  &  ENSG00000070061.16 &       1463 &     \colorbox{cyan}{ELP1} \\
6  &  ENSG00000139624.14 &       1350 &    \colorbox{cyan}{CERS5} \\
7  &  ENSG00000077254.14 &       1347 &    \colorbox{cyan}{USP33} \\
8 &  ENSG00000138614.15 &       1249 &   INTS14 \\
9 &  ENSG00000090054.15 &       1202 &   \colorbox{cyan}{SPTLC1} \\
10  &  ENSG00000183036.11 &       1187 &     \colorbox{cyan}{PCP4} \\
11  &  ENSG00000130717.13 &       1177 &     UCK1 \\
12  &  ENSG00000184164.15 &       1157 &   \colorbox{cyan}{CRELD2} \\
13  &  ENSG00000125633.11 &       1138 &   CCDC93 \\
14  &  ENSG00000128829.12 &       1112 &  EIF2AK4 \\
\hline
\end{tabular}}
    \caption{Key genes in the WSI/RNA-seq matching process. Highlighted genes have been found in glioblastoma literature.}
    \label{tab:genes}
\end{table}

\section{Discussion}
\label{sec:disc}

The highlighted genes have already been identified in literature as playing an important role in the occurrence or development of glioblastoma \cite{SGMS1,WDR11a,WDR11b,FAP,USP33a} (most of them playing a role in metabolism, micro-environment and invasion), other brain cancer types \cite{ELP1} and cancer in general \cite{GPR161,CERS5,USP33b,SPTLC1,PCP4a,PCP4b,CRELD2}.\\
Several genes identified in our study are involved in sphingomyelin metabolism. Notably, \textbf{Sphingomyelin Synthase 1 (SMS1)}, encoded by the \textbf{SGMS1} gene, is frequently downregulated in various solid cancers, including melanoma, where low SGMS1 expression has been associated with poor prognosis in metastatic cases \cite{bilal2019sphingomyelin}. Conversely, high SMS1 mRNA levels have been linked to favorable prognosis in glioma. Furthermore, the pharmacological effect of \textbf{2-hydroxyoleic acid (2OHOA)}—currently under clinical investigation for glioblastoma treatment—is attributed to its ability to upregulate SMS1 expression and enzymatic activity, while simultaneously downregulating SMS2 in glioma cells, thereby modulating cell proliferation and patient prognosis \cite{fernandez2019opposing,barcelo2011sphingomyelin}.\\
\textbf{CERS5}, responsible for C16-ceramide synthesis, plays a key role in sphingolipid signaling, tumor growth, and apoptosis. CERS5 knockout has been shown to reduce glioma stem cell proliferation and sphere formation, suppress brain xenograft growth, and extend animal survival \cite{song2024rna}. \\
\textbf{GNPTAB}, which encodes N-acetylglucosamine-1-phosphotransferase, is essential for mannose-6-phosphate (M6P) tag formation and is involved in cholesterol metabolism, lysosomal function, and autophagy regulation \cite{mareninova2021dysregulation}.\\
\textbf{SPTLC1}, encoding the long-chain base subunit of serine palmitoyltransferase, catalyzes the first step in sphingolipid synthesis and has been implicated in multiple cancers. Decreased SPTLC1 expression was observed in renal cell carcinoma (RCC), correlating with poor survival. Forced expression of SPTLC1 inhibited cell growth in vitro and in vivo, partly via modulation of the Akt/FOXO1 signaling pathway \cite{kong2019sptlc1}.\\
Our study also highlights several \textbf{oncogene regulators and tumor suppressor genes}. For example, \textbf{GPR161}, a cilium-localized GPCR, suppresses Sonic hedgehog (Shh) signaling in the neural tube \cite{shimada2018basal}. \textbf{ELP1} and \textbf{WDR11} have both been implicated in the pathogenesis of medulloblastoma and glioblastoma, respectively, through loss-of-function mutations \cite{niu2020five,wei2017exome}. Conversely, \textbf{INTS14} has been shown to activate MYC promoter activity. Knockdown of INTS14 in prostate cancer cell lines led to reduced MYC expression and G0/G1 cell cycle arrest \cite{tanaka2022efficient}.\\
\textbf{Dysregulated endocytic trafficking} of oncogenic receptors, such as members of the EGFR family, contributes to persistent oncogenic signaling. \\ \textbf{COMMD3}, a key regulator of HER2 endosomal trafficking, significantly inhibits proliferation, migration, and epithelial-mesenchymal transition (EMT) in ovarian cancer cells, and suppresses tumor invasion in murine models \cite{wang2023commd3}. In glioblastoma, where EGFR is frequently mutated or overexpressed, COMMD3 may similarly regulate tumor invasion and EMT.\\
\textbf{Ferroptosis}, an autophagy-dependent form of cell death, has been shown to regulate glioma stem cell death. \textbf{EIF2AK4 (GCN2)} is a kinase involved in this pathway, phosphorylating eIF2$\alpha$ to block translation initiation. EIF2AK4 expression was found to be elevated specifically in glioblastoma tissues compared to non-tumoral brain tissues, suggesting a role in tumor cell adaptation \cite{marina2019relevance}.\\
\textbf{Tumor microenvironment} plays a critical role in glioblastoma aggressiveness and therapy resistance. This includes angiogenesis, hypoxia, stromal fibroblasts, and immune cells such as macrophages. Our study identified three genes involved in microenvironment regulation: \textbf{FAP, USP33} and \textbf{CRELD2}.
\begin{itemize}
    \item \textbf{FAP (Fibroblast Activation Protein} is expressed in tumor vessels and neoplastic cells and is characteristic of mesenchymal cells, including cancer-associated fibroblasts and pericytes. FAP+ pericyte-like cells are major producers of extracellular matrix components like collagen I and fibronectin 1 in GBM, which promote glioma cell migration via FAK activation \cite{vymola2024fibrillar}. Additionally, FAP-targeting CAR-T cells were shown to effectively control tumor growth in models containing both antigen-positive and -negative glioblastoma cells \cite{yu2024endogenous}.
    \item \textbf{USP33} is a deubiquitinase that stabilizes HIF-2$\alpha$ in an ERK1/2-dependent manner, enhancing hypoxia responses in glioma stem cells. Its knockdown impairs stem cell maintenance and tumor vascularization, reducing glioblastoma growth \cite{zhang2022usp33}.
    \item \textbf{CRELD2} is an ER-stress responsive gene implicated in microenvironment regulation, particularly in breast cancer. \cite{boyle2020rock} demonstrated that CRELD2 acts as a paracrine factor downstream of the ROCK-PERK-ATF4 signaling axis, promoting fibroblast education and tumor progression \cite{tang2023creld2}.
\end{itemize}

\section{Conclusion}
In conclusion, our study demonstrates the effectiveness of multimodal deep learning approaches in identifying genetic profiles that explain different glioblastoma patterns. By leveraging joint image/RNA-seq analysis and introducing novel methods to encode RNA-seq data, we have shed new light on the heterogeneous nature of this aggressive brain tumor. Our findings not only confirm existing medical literature but also highlight new potential targets for therapeutic intervention. These results have significant implications for the development of personalized medicine strategies for glioblastoma patients and underscore the importance of continued research into the application of AI algorithms in cancer biology.

\section*{Acknowledgement}
The results shown here are in part based upon data generated by the TCGA Research Network: \url{https://www.cancer.gov/tcga}. The authors would like to thank Caroline Delmas and Antony Lemarié\footnote{INSERM UMR 1037 Cancer Research Center of Toulouse (CRCT), University Paul Sabatier Toulouse III} for WSI and RNA-seq data acquisition.


\newpage
\appendix
\section{Hyperparameters \& training details}
\paragraph{Hardware requirements}
All training and data preparation was done using a single machine with the following characteristics:
\begin{itemize}
    \item CPU: AMD Ryzen Threadripper 3970X 32-Cores
    \item RAM: 128 GB DDR4
    \item GPU: 2x RTX3090 (24GB of VRAM)
    \item OS: Ubuntu 24.04 LTS
\end{itemize}

Training the RNA-seq encoder requires 8 minutes, training the WSI encoder requires 23 hours and training the multimodal model with no frozen parts, requires 9 hours.

\begin{figure}[htbp]
    \centering
    \includegraphics[width=0.99\linewidth]{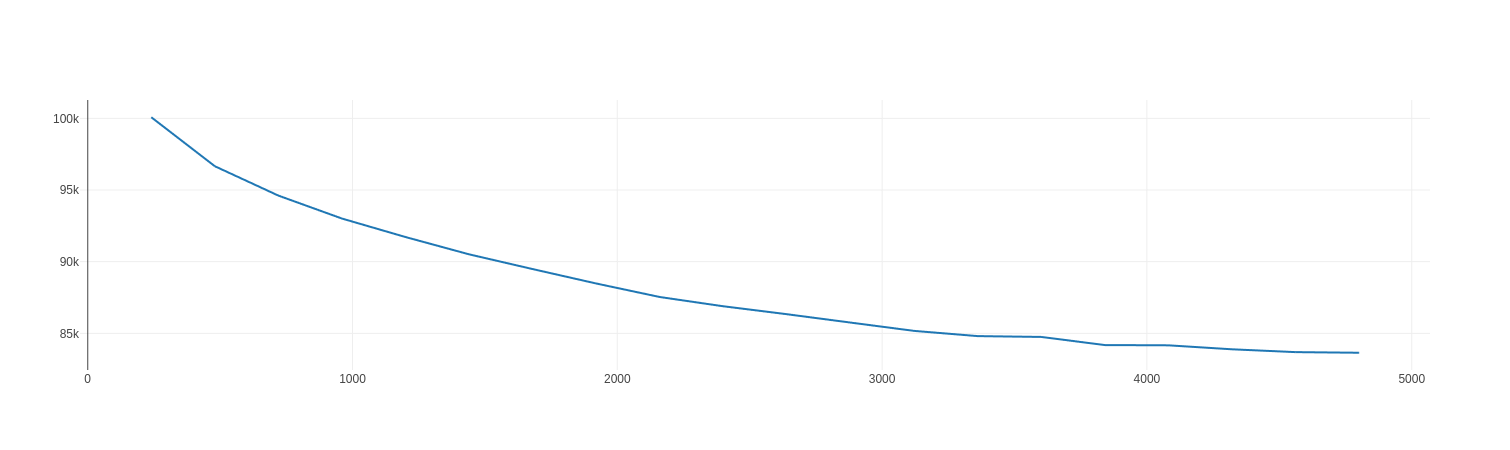}
    \includegraphics[width=0.99\linewidth]{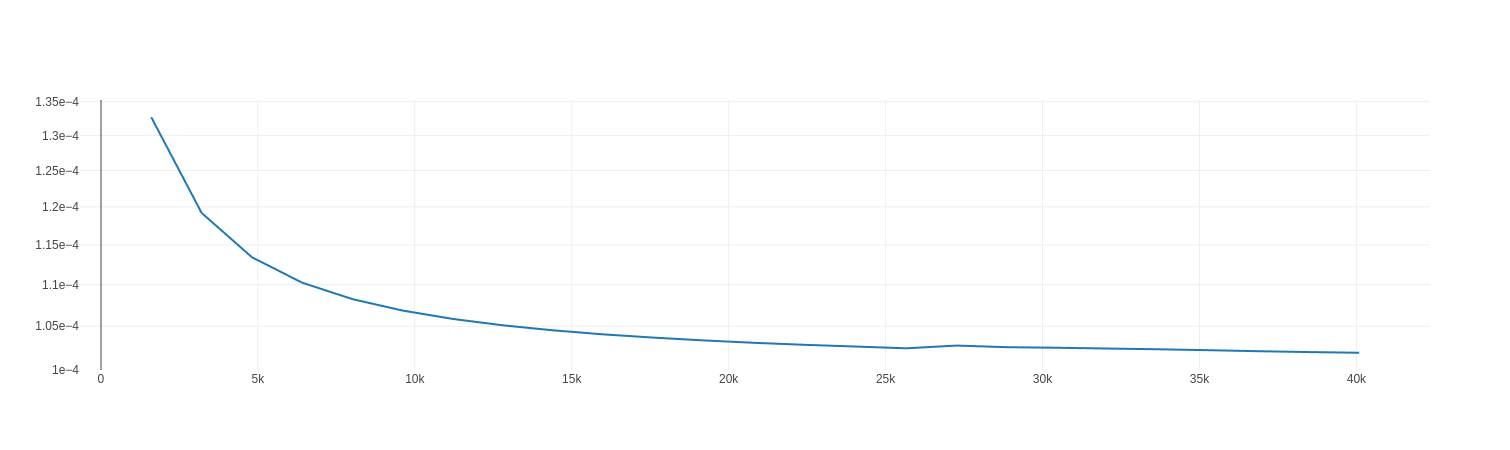}
    \includegraphics[width=0.99\linewidth]{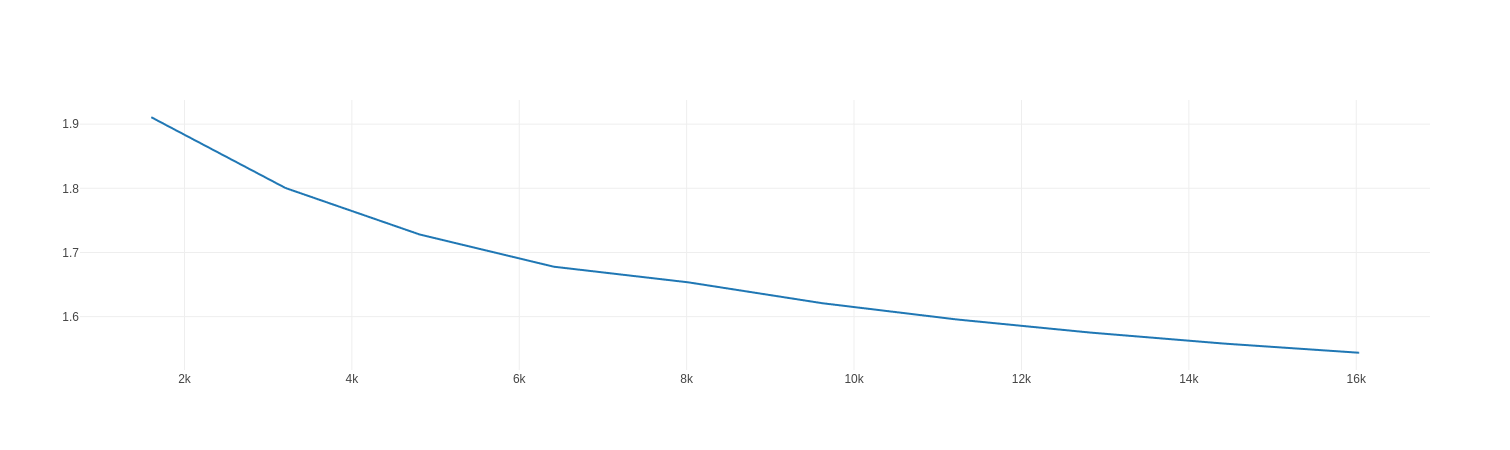}
    \caption{Validation losses w.r.t training steps for RNA-seq encoder pre-training (upper), WSI encoder pre-training (center) and multimodal model training (lower).}
    \label{fig:plots}
\end{figure}

\paragraph{Training and validation curves}
Fig.\ref{fig:plots} provides the plots of losses after each epoch on the validation data for the RNA-seq encoder, WSI encoder and multimodal model. All plots show no sign of model overfitting. Training was stopped when validation performance did not increase after five consecutive training epochs.

\paragraph{Hyperparameters}
The following tables provide the main hyperparameters for each training step (RNA-seq encoder pre-training, WSI encoder pre-training and multimodal training). These hyperparameters led to the best performance figures, as indicated in table \ref{tab:results}.

\begin{table}[htbp]
\begin{minipage}[b]{0.49\textwidth}
\centering
\resizebox{\textwidth}{!}{
\begin{tabular}{|c|l|c|}
\hline
\textbf{Parameter} & \textbf{Name} & \textbf{Value} \\ \hline\hline
\multirow{4}{*}{\textbf{Model}} 
& Embedding dimension & 128 \\
& Dropout probability & 0.5 \\
& Number of heads & 8 \\
& Projection dimension &  64 \\ \hline
\multirow{3}{*}{\textbf{Optimizer}}  & \textbf{AdamW} & \\
& Learning rate & $1 \times 10^{-4}$ \\
& Weight decay & 0.02 \\ \hline
\multirow{6}{*}{\textbf{Scheduler}} & \textbf{Cosine scheduler} & \\
& Epochs & 20 \\
& Minimum learning rate & $1 \times 10^{-5}$ \\
& Decay rate & 1 \\
& Warmup learning rate & $1 \times 10^{-5}$ \\
& Warmup epochs & 6 \\ \hline
\multirow{6}{*}{\textbf{Training}} & Batch size & 32 \\
& Maximum epochs & 20 \\
& Temperature & 0.005 \\
& Mask probability & 0.30 \\
& Coefficient reconstruction loss & 1.0 \\
& Coefficient contrastive loss & 1.0 \\ \hline
\end{tabular}
}
\end{minipage}
\begin{minipage}{0.01\textwidth}
\end{minipage}%
\begin{minipage}[b]{0.49\textwidth}
\centering
\resizebox{\textwidth}{!}{
\begin{tabular}{|c|l|c|}
\hline
\textbf{Parameter} & \textbf{Name} & \textbf{Value} \\ \hline\hline
\multirow{4}{*}{\textbf{Model}} 
& Embedding dimension & 256 \\
& Dropout probability & 0 \\
& Number of heads & 12 \\
& Mini-patch size &  $16\times16$ \\ \hline
\multirow{3}{*}{\textbf{Optimizer}}  & \textbf{AdamW} & \\
& Learning rate & $1 \times 10^{-4}$ \\
& Weight decay & 0.02 \\ \hline
\multirow{6}{*}{\textbf{Scheduler}} & \textbf{Cosine scheduler} & \\
& Epochs & 25 \\
& Minimum learning rate & $1 \times 10^{-5}$ \\
& Decay rate & 1 \\
& Warmup learning rate & $1 \times 10^{-5}$ \\
& Warmup epochs & 6 \\ \hline
\multirow{3}{*}{\textbf{Training}} & Batch size & 64 \\
& Maximum epochs & 25 \\
& Margin & $1\times 10^{-4}$ \\ \hline
\end{tabular}
}
\end{minipage}
\caption{Hyperparameters for RNA-seq (left) and WSI (right) encoders pre-training}
\label{tab:pre-hyperparameters}
\end{table}

\begin{table}[htbp]
\centering
\begin{tabular}{|c|l|c|}
\hline
\textbf{Parameter} & \textbf{Name} & \textbf{Value} \\ \hline\hline
\multirow{4}{*}{\textbf{Model}} 
& Co-attention dimension & 768 \\
& Number of heads & 8 \\
& Frozen RNA-seq encoder & \textit{False} \\
& Frozen WSI encoder &  \textit{False}\\ \hline
\multirow{3}{*}{\textbf{Optimizer}}  & \textbf{AdamW} & \\
& Learning rate & $1 \times 10^{-4}$ \\
& Weight decay & 0.02 \\ \hline
\multirow{6}{*}{\textbf{Scheduler}} & \textbf{Cosine scheduler} & \\
& Epochs & 10 \\
& Minimum learning rate & $1 \times 10^{-5}$ \\
& Decay rate & 1 \\
& Warmup learning rate & $1 \times 10^{-5}$ \\
& Warmup epochs & 6 \\ \hline
\multirow{5}{*}{\textbf{Training}} & Batch size & 64 \\
& Maximum epochs & 10 \\
& Coefficient reconstruction loss & $0$ \\
& Coefficient contrastive loss & $1$ \\
& Coefficient matching loss & $1$ \\ \hline
\end{tabular}
\caption{Hyperparameters for multimodal model training}
\label{tab:multimodal-hyperparameters}
\end{table}

 \newpage
 \bibliographystyle{elsarticle-num} 
 \bibliography{cas-refs_v2}

\end{document}